\documentclass[11pt]{article}
\usepackage{moriond,epsfig}

\def\be{\begin{equation}}
\def\ee{\end{equation}}
\def\bea{\begin{eqnarray}}
\def\eea{\end{eqnarray}}

\newcommand{\Et}{\mbox{$E_T$}}

\newcommand{\met}{\mbox{$\protect \raisebox{0.3ex}{$\not$}\Et$}}

\begin{document}
\vspace*{4cm}
\title{Top quark properties at the Tevatron}

\author{Fabrizio Margaroli}

\address{Purdue University, Department of Physics, 525 Northwestern Avenue,\\
West Lafayette, Indiana 47907, USA}

\maketitle\abstracts{
The discovery of the top quark in 1995 opened a whole new sector of investigation of the Standard Model; today top quark physics remains a key priority of the Tevatron program. Some of the measurements of top quark properties, for example its mass, will be a long-standing legacy. 
The recent evidence of an anomalously large charge asymmetry in top quark events suggests that new physics could couple preferably with top quarks. I will summarize this long chapter of particle physics history and discuss the road the top quark is highlighting for the LHC program.
}

\section{Introduction}

The top quark is the last discovered quark, and by far the most massive particle in the Standard Model (SM) of particle interactions, its mass being approximately 30 times larger than the next-to-heaviest quark. The value close to 175\,GeV$c^2$ surprised the high energy community as it corresponds to a Yukawa coupling of the top quarks very close to one, suggesting a possible special role of the top quark in the electroweak symmetry breaking mechanism. Another very interesting consequence of its very large mass is the top quark lifetime being shorter than the hadronization time: $1/\Gamma_{top} < 1/\Lambda_{QCD}$: the top quark is the only quark that decays before hadronizing, and thus the only quark that can be studied naked. 
According to the SM, top quarks decay almost always to a $W$ boson and a $b$ quark.
Top quarks can be produced singly through electroweak interactions, or more abundantly in pairs through QCD interactions. The pair production cross section is approximately 7.5\,pb, more than a factor of two larger than the single top quark production cross section. More importantly, the larger energy scale of the pair production process and the striking $W^+W^- b \bar b$ decays allow for a much better background rejection than in the single top quark production, thus making the pair production the favored channel for top quark properties studies. The signature of $t \bar t \to Wb Wb$ events is completely characterized by the leptonic or hadronic decays of the two $W$ bosons. It is tradition to call ``dileptonic decays" the events where both $W$ bosons decay leptonically, ``semileptonic" decays the ones where only one $W$ boson decays leptonically, and ``all-hadronic decays" the ones where both $W$ bosons decay hadronically. The branching ratios (BR) are 10\% 44\% and 46\% respectively. In the dilepton channel, the mere requirement of two leptons, missing transverse energy ($\met$) as a signature of neutrinos, and jets gives a good signal-to-background (S/B) ratio. In the semileptonic channel, it is common to add to the lepton, $\met$ and jets requirement, the additional requirement of at least one jet being identified as originating from a $b$ quark ($b$-tagged jet) to achieve a good S/B ratio. In the all-hadronic channel, the requirement of at least one $b$-tagged jet and several high $p_T$ jets in the final state is not sufficient to suppress the overwhelming QCD multijet background - additional kinematical and topological requirements are needed. The ideal balance between the branching ratio, S/B, and overall event reconstruction is achieved in the semileptonic channel, which is thus the best suited for most top quark properties measurements. Having collected more than 8.5\,fb$^{-1}$ of integrated luminosity at both experiments as of this conference, several top quark properties have been measured with increasing precision, while other properties have been investigated for the first time.  
\section{Intrinsic top quark properties}
The CDF and D0 collaborations have measured the width of the top quark, excluded exotic values for its charge, and measured its mass. Among the intrinsic top quark parameters, the mass is the one that deserves most attention. In fact, the top quark mass is a free parameter in the SM, and indirect determination using constraints from precisely known electroweak parameters\,\cite{GFitter} can predict its value with limited precision. In turn, the precision on the top quark mass measurements is of crucial importance as the top quark mass is the single most sensitive parameter once using electroweak data to constrain the SM Higgs boson mass range\,\cite{GFitter,thisconf}. The same experimental set of inputs, top quark mass included, is used to constrain a large number of new physics scenarios: extra dimensions, technicolor, fourth fermion generation, etc.
From an experimental point of view, the precise knowledge of the top quark mass reflects in more precise predictions on the top quark production cross sections at colliders, and understanding of the production and decay kinematics. Thus the precise measurement of the top quark mass translates into refined measurements of all top quark properties.

\begin{figure}
\psfig{figure=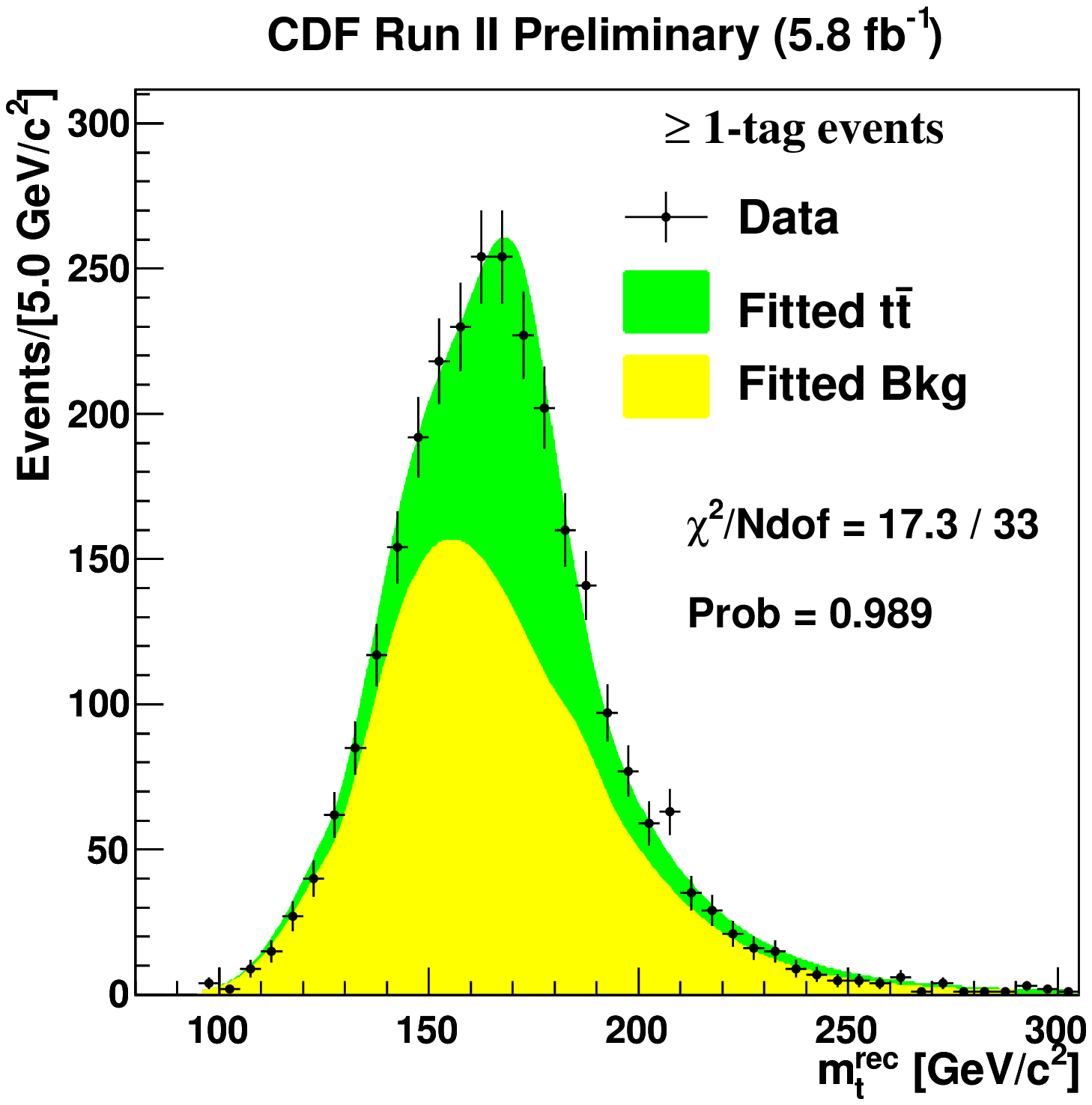,height=2.9in}
\psfig{figure=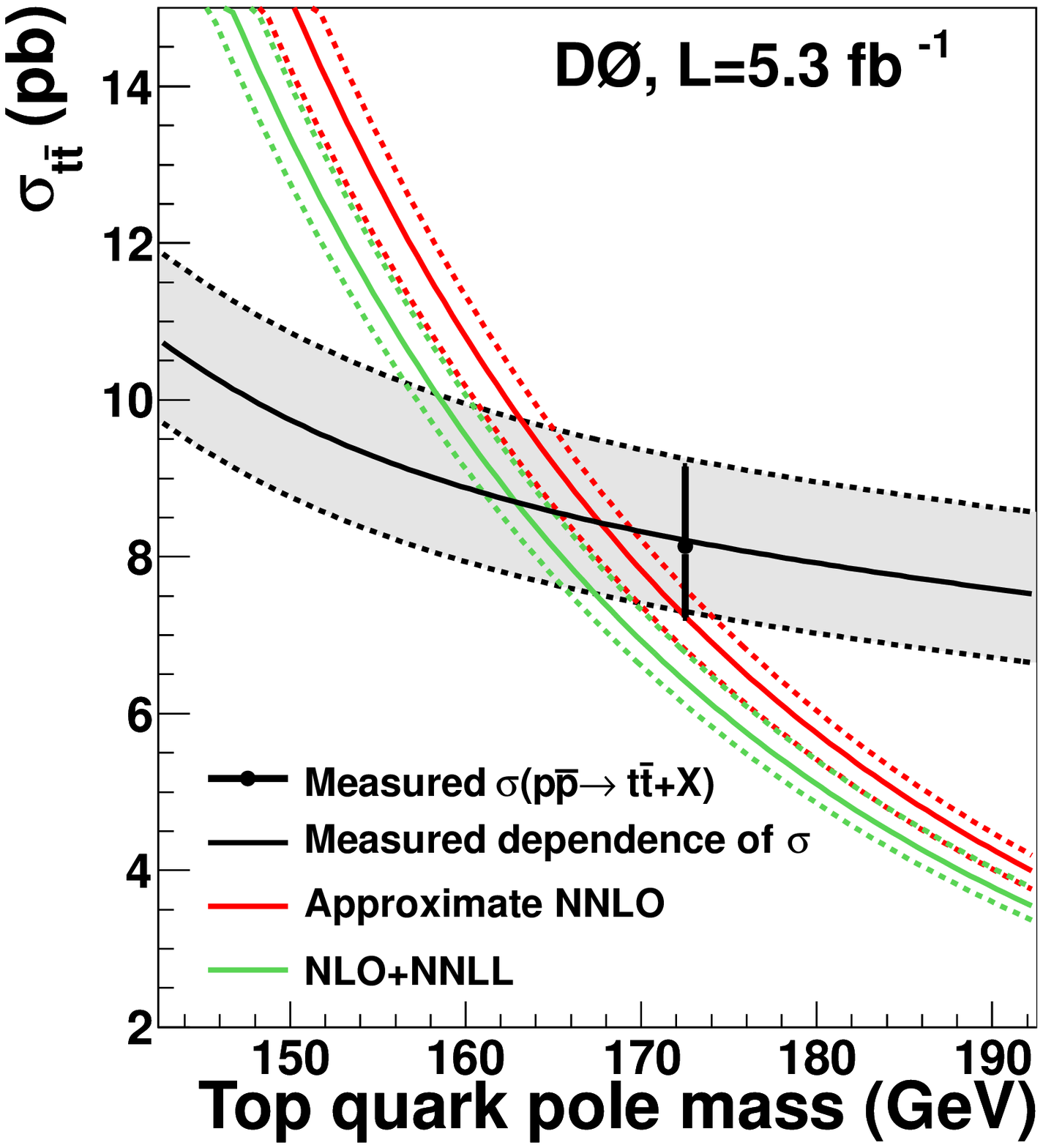,height=2.9in}
\caption{The left plot shows the reconstructed top quark invariant mass distribution in the all-hadronic decay mode, for events passing kinematical and topological cuts, and the requirement of at least one $b$-tagged jet. The green area represent the signal contribution as modeled with a Monte Carlo simulation of top quark mass equal to 172.5\,GeV/$c^2$. The right plot shows the top quark mass extracted from the D0 cross section measurement in the lepton+jets channel, in the hypothesis that the Monte Carlo encoded mass is the top quark pole mass.}
\label{fig:Mtop}
\end{figure}

As mention in the Introduction, the semileptonic channel is the one that provides the best sensitivity to top quark mass measurements. Still, different channels provide statistically independent results, and are affected by different systematics; it is thus of the utmost importance to measure the top quark mass in all available channels. The jet energy scale (JES) uncertainty is the largest systematic affecting top quark mass measurements, and it has been reduced sensibly by calibrating it {\em in situ} using events where at least one $W$ boson decays hadronically. In this regard, the all-hadronic decays have the advantage of having two $W$ bosons to calibrate the top quark mass. CDF updated a previously published measurement\,\cite{Aaltonen:2010pe} using twice the data, totaling 5.8\,fb$^{-1}$. The event selections requires 6-to-8 jets in the final state, at least one $b-$tagged jet, low $\met$, and utilizes a neural network to optimally suppress the QCD background while retaining large acceptance to the signal. The main challenge has been to cope with the increasing Tevatron instantaneous luminosity and the consequently increasingly larger backgrounds; CDF dealt with it by cutting to higher neural-network output value. CDF measures $M_{top} = 172.5 \pm 1.7 (stat+JES) \pm 1.2 (syst)$GeV/$c^2$, where the first uncertainty contains the statistical part of the JES uncertainty as it is measured {\em in situ}. The distribution of the reconstructed top quark mass for the signal plus background can be seen in Figure\,\ref{fig:Mtop}.

Due to limitation in tracking and muon chamber pseudorapidity coverage, a relatively large fraction of forward electrons and muons are not identified at CDF. Also, the identification of hadronically decaying taus is especially difficult. A new CDF measurements measures the top quark mass in events selected with large $\met$, at least four jets where at least one is $b$-tagged, and further applies topological and kinematical requirements through a neural network, to recover the semileptonic events where electrons/muons/taus are not identified. The sample composition has been understood in the context of an earlier measurement of $\sigma_{t \bar t}$ in the same dataset\,\cite{Aaltonen:2011tm}. In these events, it is impossible to fully reconstruct the kinematics due to the undetected $W \to \ell \nu_{\ell}$. Still, the reconstruction of only one of the two top quarks is sufficient to extract an $M_{top}$ measurement. Using 5.6\,fb$^{-1}$ of data, CDF measures\,\cite{CDFmetjets} $M_{top} = 172.3 \pm 2.4 (stat+JES) \pm 1.0 (syst)$\,GeV/$c^2$. To increase precision in the $M_{top}$ determination, CDF combines the most precise measurement in each decay mode. The event selection for the several decay channels are devised to select independent datasets. The combination requires properly taking into account the correlation of the systematic uncertainties among all measurements. CDF uses the BLUE method\,\cite{BLUE} and obtains an average of $M_{top} = 172.7 \pm 0.6 (stat) \pm 0.9 (syst)$\,GeV/$c^2$. The CDF average shown at this conference is as precise as the 2010 Tevatron average\,\cite{:1900yx}. All the measurements that enter into the CDF, D0, and thus the Tevatron combination are calibrated to Monte Carlo simulations. Thus the ``mass" measured is effectively the definition contained in the leading order (LO) Monte Carlo used; theorists agree that the Monte Carlo mass should be very close to the top quark pole mass\,\cite{Glenzinski:2008zz}.
Beyond LO quantum chromodynamics (QCD), the mass of the top quark is a convention-dependent parameter, the other dominant convention being the MS scheme. To probe further into this ambiguity, D0 compares the measured inclusive $t \bar t$ production cross section\,\cite{Abazov:2011mi} with fully inclusive calculations at higher-order QCD that involve an unambiguous definition of $M_{top}$\,\cite{MochUwer,Ahrens} and compares the results to MC\,\cite{Abazov:2011pt}. The measurement favors the pole mass hypothesis over the $\mathrm{\bar MS}$ hypothesis. 

By the end of the Tevatron run, the CDF and D0 collaborations will analyze a dataset more than twice as large as the one shown at this conference. It is expected that the precision on $M_{top}$ measurements will drop below 1\,GeV/$c^2$, and that the scheme interpretation would be better understood.
\section{Top quark production properties}

\begin{figure}
\psfig{figure=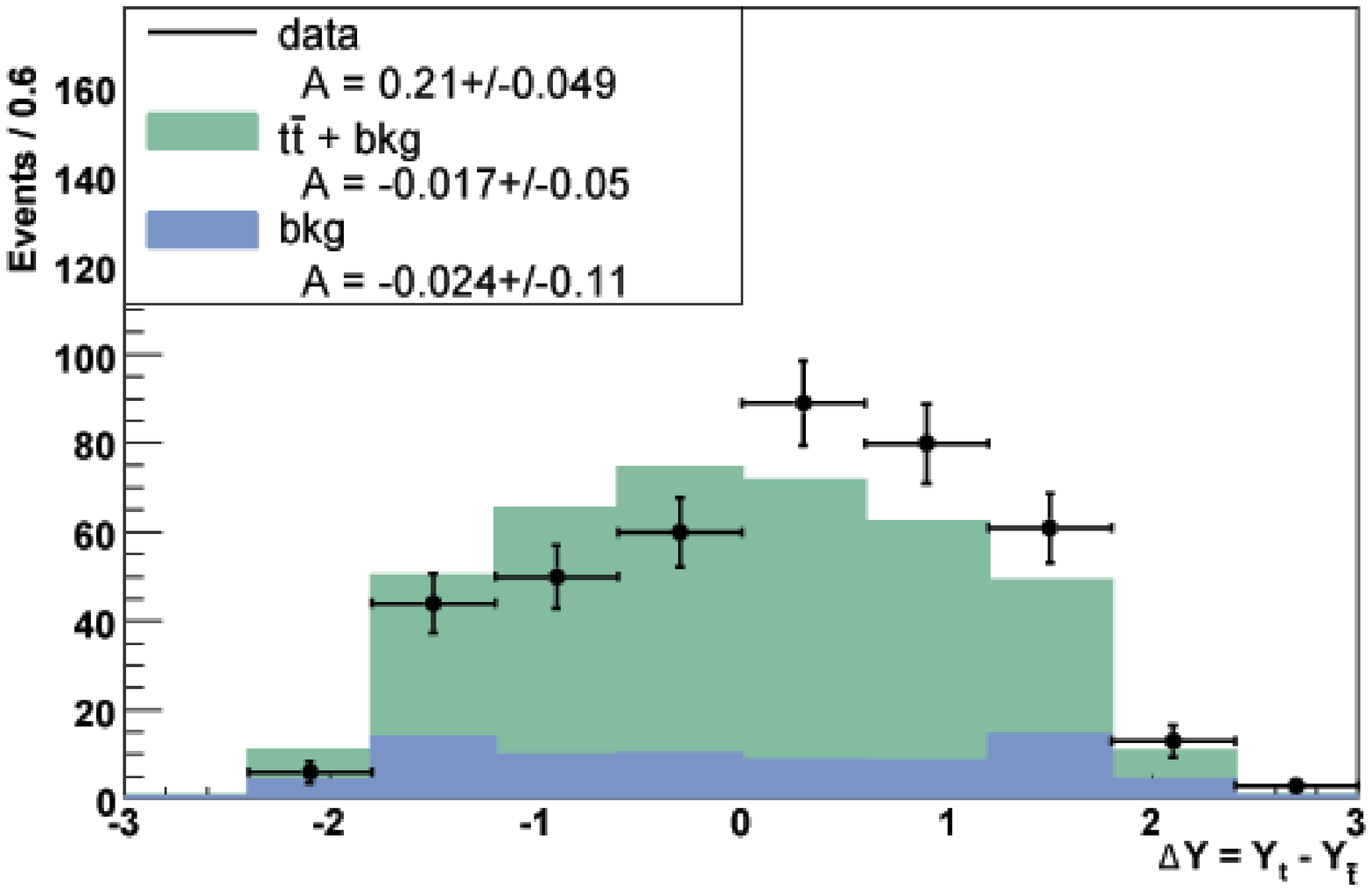,height=1.95in}
\psfig{figure=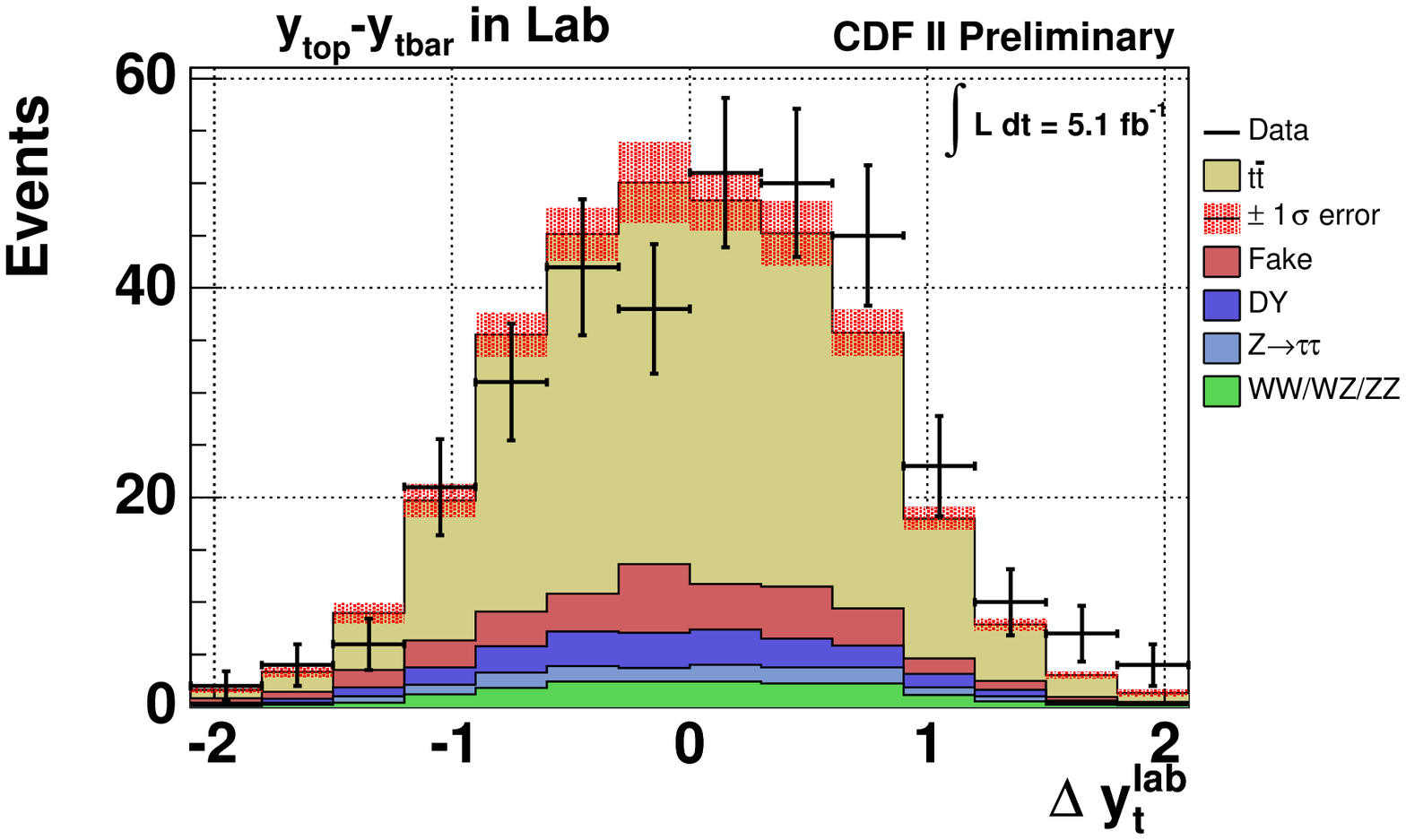,height=2.05in}
\caption{The left plot shows the distribution of $\Delta y$ at high top-antitop invariant mass in the lepton+jets channel.
The right plot shows the $\Delta y$ distribution in the dilepton channel. No invariant mass cut is applied in the latter plot.
\label{fig:AFB}}
\end{figure}

The top quark pair production cross section has been measured in all decay modes by the Tevatron (and in the semileptonic and dileptonic modes at the LHC collider). The most precise determination comes from CDF\,\cite{thisconf2} and has a precision of $6.5\%$ and is in excellent agreement with approximate NNLO\,\cite{MochUwer} QCD computation and consistent with NLO+next-to-leading-log (NLL) computation\,\cite{Ahrens}. While these results are impressive, there is still room for a contribution from new physics at top quark production level of the order of 0.5 - 1.0\,pb. A sensitive probe of the SM nature of top quark production at the Tevatron $p \bar p$ collider is the measurement of the forward-backward asymmetry of top quarks production, where ``forward" (``backward") stands for ``along the proton (antiproton) direction". QCD predicts a positive asymmetry of about 5\% originating from interference effects between LO and NLO diagrams\,\cite{Campbell:2010ff}. Early CDF and D0\,\cite{D0Afb} measurements in the semileptonic decay mode showed a positive deviation of approximately 2\,$\sigma$ from the SM predictions. Several new physics scenarios predict this asymmetry to be enhanced due to $t \bar t$ creation through exotic mechanisms contributing to $s-$ or $t-$channel diagrams\footnote{One notable exception would be the possibility of a new process giving rise to top quarks and invisible particles that blends with the SM $t \bar t$ production\,\cite{Isidori:2011dp}.}. In most scenarios, the departure from SM prediction increases by looking at particular regions of the phase space. Using about 5\,fb$^{-1}$ of data, CDF investigated this asymmetry as a function of the rapidity difference among top and antitop quarks, and as a function of the top-antitop system invariant mass\,\cite{Aaltonen:2011kc}. The asymmetry was found to be the largest for events with $M(t \bar t) > 450$\,GeV/$c^2$, amounting to $48 \pm 11\%$, more than 3\,$\sigma$ deviation from the SM predictions. An independent measurement performed in the dileptonic decay mode observes an inclusive asymmetry of $42 \pm 16\%$, corresponding to a 2.3\,$\sigma$ deviation from the SM\,\cite{CDFAfbdil}. The rapidity difference between the top and antitop quarks in the semileptonic and dileptonic decay modes can be observed in Figure\,\ref{fig:AFB}. The statistics in the dileptonic decay mode is not yet sufficient to establish a trend of the asymmetry as a function of the total invariant mass. 
In summary, the CDF and D0 collaborations observe a consistently larger-than-expected asymmetry. CDF notes that this asymmetry (and the deviation from the SM) grows with the invariant mass of the system. This latter result ignited the already present interest by theoreticians, and is currently being used to set constraints to new physics scenarios affecting top quark production. One particular model, the $t-$channel production through leptophobic $Z^{\prime}$ boson exchange, saw a surge in popularity after CDF published a study showing a possible new resonance decaying to two jets in the $\ell \nu jj$ data sample\,\cite{Aaltonen:2011mk}. More studies are ongoing at both collaborations.

At the Tevatron, the $p \bar p$ collision center-of-mass (COM) energy of 1.96\,TeV is large enough for heavy particles to be produced occasionally with very large Lorentz boost. The identification of boosted heavy objects decaying to jets will be a very useful tool in the search for new physics at colliders with even larger COM such as the LHC, as it would provide large discrimination of interesting signals against the much more common QCD background. One notable example would be the search for the low mass Higgs boson in $\ell \nu b \bar b$ events, where with sufficient boost the $b \bar b$ system would appear as a single massive jet. Another example is the search for heavy resonances decaying to top quarks, as for example a leptophobic $Z^{\prime}$ or an axigluon of the kind that could explain the anomalously large forward backward asymmetry described in this Section. CDF performed the first search for SM $t \bar t$ production through the boosted top signature\,\cite{Boost}. The SM predicts only a very small fraction of top quarks to be produced with $p_T^{top} >400GeV$, corresponding to a cross section of approximately a few fb. In this momentum range, the top daughter decay particles would appear as highly collimated. In case of an hadronic top decay $t \to W b \to qqb$ the signature would be that of a large jet with substructure. The decays of a boosted top with decays $t \to Wb \to \ell \nu b$ are more difficult as leptons are usually required to be isolated to suppress the fake rate. CDF looks at both the all-hadronic (semileptonic) final state, reconstructing jets using a clustering algorithm with the cone size parameter set to 1.0, significantly larger than the one used in top quark physics. The jets mass (and missing transverse energy significance) are used as discriminants against the vast background of QCD production of lighter quarks and gluons. In the case of boosted top quarks decaying hadronically, the jet mass is in fact close to the top quark mass itself, while it peaks at lower values in the case of QCD light quark/gluon production. For top quarks decaying to $\ell \nu b$, the requirement of large missing trasnverse energy alone is used to reject the QCD background. Analyzing 6\,fb$^{-1}$ of data, CDF finds a modest excess of events: 58 candidate events with an estimated background of $44 \pm 15$ events. In absence of a signal, this analysis sets a 95\% confidence level (C.L.) upper limit on the rate of top quark production for top quarks with $p_T > 400$\,GeV/$c$ of 40\,fb. The same data is used to search for pair production of a massive object with masses comparable to that of the top quark, setting an upper limit on the pair production of 20\,fb at 95\% C.L..

\section{Top quark decay properties}

While top and anti-top quarks are pair produced unpolarized at hadron colliders, the orientations of their spins are correlated. Top quarks are the ideal laboratory for spin correlation studies as they decay before this correlation can be affected by the fragmentation process. Spin correlation is also a sensitive probe of beyond the standard model scenarios, as different model would predict different top polarizations. The charged leptons from the $t \to W b \to \ell \nu b$ decays are the probes with the highest sensitivity to the direction of the top quark spin.
In a recent paper\,\cite{Abazov:2011qu} D0 measures the strength of the $t \bar t$ spin correlation $C$ from a differential angular distribution involving the angles between the flight direction of the two decay leptons in the rest frames of their respective top quarks and the spin quantization axis. NLO QCD computation predicts this quantity to be $C=0.78\pm0.03$. The measurement uses 5.4\,fb$^{-1}$ of $p \bar p$ collision and finds $C=0.10\pm0.45$, in reasonable agreement with the SM predictions.

\begin{figure}
\psfig{figure=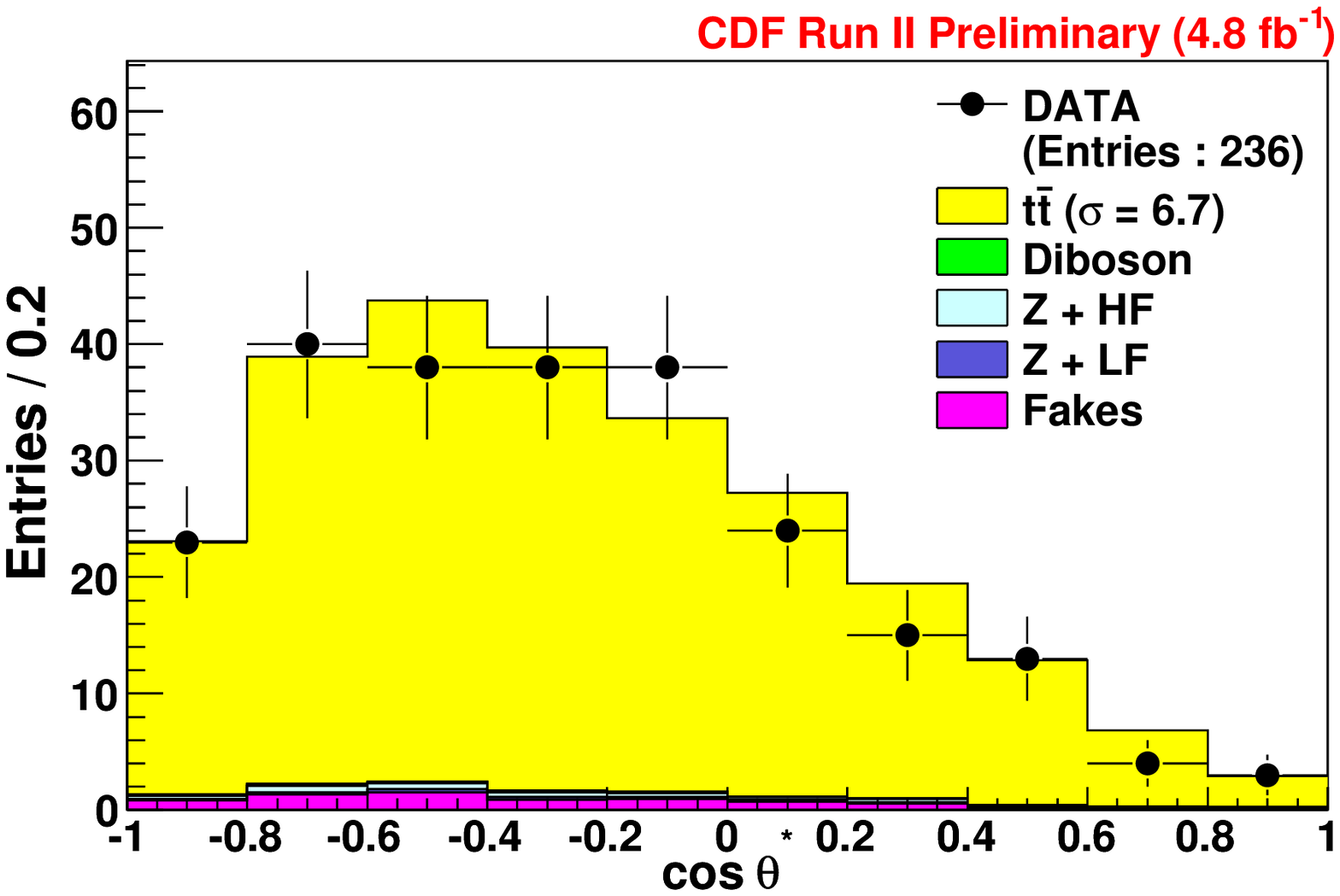,height=2.3in}
\psfig{figure=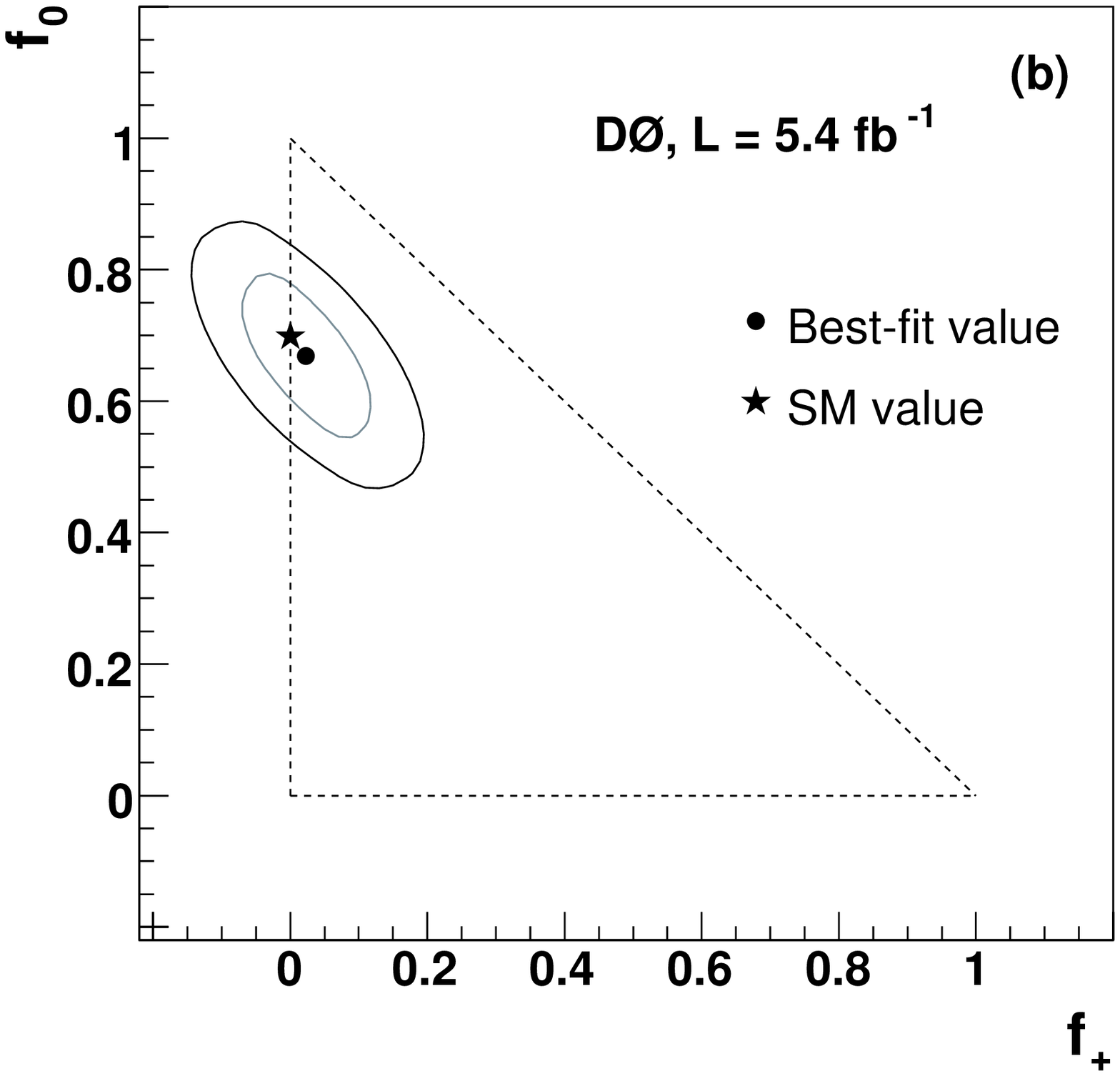,height=2.65in}
\caption{The left plot shows the $cos\theta^*$ distribution comparison between data and expected SM contributions in the CDF dilepton sample. The right plots shows the model-independent D0 measurement in the 2-dimensional plane of f$_+$ and f$_0$, along with the SM predictions.
\label{fig:whel}}
\end{figure}

The on-shell W bosons from top quark decays can have three possible helicity states. SM predicts that the top quark decays via the $V - A$ charged weak current interaction, which strongly suppresses the presence of right-handed W bosons. The fraction of $t \bar t$ events where the $W$ bosons are produced longitudinally polarized, $f_0$, left-handed, $f_-$ or right-handed, $f_+$ depend upon the masses of the top quark $M_{top}$ and $W$ boson ($M_W$). Using the current world average values for $M_{top}$ and $M_W$,  the expected SM values are\,\cite{WhelTheory}. $f_0=0.698$, $f_-=0.301$, and $f_+ = 4.1 \times 10^{-4}$. A significant deviation from these predictions would provide strong evidence of physics beyond the SM, indicating either a departure from the expected $V - A$ structure of the $t W b$ vertex, or the presence of a non-SM component in the $t \bar t$ candidate sample. CDF and D0 measured simultaneously the fraction of longitudinal and right-handed $W$ bosons from top quark decays in the semileptonic and dileptonic decay modes. The latest CDF measurement\,\cite{CDFWhelDil} analyzes 4.8\,fb$^{-1}$ of dileptonic events to measure the following fractions $f_+=0.12\pm0.12$ and $f_0=0.78\pm0.20$. In order to measure the $W$ boson helicity fractions, the angle $\theta^*$ between the down-type decay product of the $W$ boson in the rest frame of the latter, with respect to the $W$ boson flight direction in the top quark rest frame is measured. The distribution of the cosine of this angle differs for the three possible helicity fractions, which is exploited in the measurement.
The plot in Figure\,\ref{fig:whel} shows the distribution for the observable sensitive to the $W$ boson polarization and the agreement with SM expectations. D0 reports a measurement in the dileptonic and semileptonic final states using 5.4\,fb$^{-1}$ of collisions\,\cite{Abazov:2010jn} of $f_+=0.02\pm0.05$ and $f_0=0.67\pm0.10$. 

The knowledge of the color-connections between jets can serve as a powerful tool for separating processes that would otherwise appear similar. For example, in the decay of a Higgs ($H$) boson to a pair of $b$ quarks, the two $b$ quarks are color connected to each other, since the $H$ is a 
color singlet,whereas in $g \to b \bar b$ background events, they are color-connected to beam remnants since the gluon is a color-octet. The technique involves measuring a vectorial quantity called Òjet pull,Ó which represents the eccentricity of the jet in the $\eta - \phi$ plane and the direction of the major axis of the ellipse formed from the jet energy pattern. Jets tend to have their pull pointing towards their color-connected partner: in 
$H \to b \bar b$ events, the pulls of the two $b$-jets tend to point towards each other, whereas in $g \to b \bar b$ events, they point in opposite directions along the collision axis. D0 tests this technique looking at the light quark jets coming from the $W$ boson decay in semileptonic events using 5.3\,fb$^{-1}$ of collisions \,\cite{Abazov:2011vh}. The data are compared to both standard model $t \bar t$ Monte Carlo (with a color-singlet W boson) and an alternative model of $t \bar t$ with a hypothetical color-octet $``W"$ boson decaying hadronically. D0 determines the fraction of events coming from color-singlet W boson decay ($f_{Singlet}$) to be $f_{Singlet} = 0.56 \pm 0.42$, in agreement with the expectation of $f_{Singlet} > 0.277$ at 95\% C.L.

Searches for flavor changing neutral current (FCNC) are also sensitive probes to new physics. The top quark decays $t \to Z u(c)$ are heavily suppressed in the SM; still, extensions of the SM such as SUSY or quark compositeness would predict much larger values. D0 looks for the first time at the possible signature of $t \bar t \to Zq Wb \to \ell^+ \ell^- q \ell b$ and in the absence of signal, sets a limit on the hypothetical branching ratio of top quarks to $Z q$: $B(t \to Zq) < 3.2\%$ at the 95\% C.L. This result surpasses previous limits\,\cite{:2008aaa}. The result presented here translates into an observed limit on the FCNC coupling of $vtqZ < 0.19$ for $M_{top} = 172.5$\,GeV/$c^2$.

\section{Conclusions}

Top quark physics has proceeded with giant's steps over the first results obtained with Tevatron Run I data. Analyzing a top quark sample 100 times larger than the one needed for its discovery, the CDF and D0 collaborations have been able to greatly expand the range of top quark properties measurements, increase precision to unprecedented levels, and to probe in a much finer detail the nature of this very peculiar quark. For 15 years, top quark physics remained a Tevatron prerogative. In 2011 the first statistically significant deviation from SM predictions appeared in the charge asymmetry of top quark pair production, immediately confirmed in an independent dataset. While the possibility of an underestimation of the SM predictions is still open, this result gave rise to en enormous interest in the theoretical community.
Twice the current dataset will be available in a few months, and it is expected that its study will increase further the precision of measurements such as the top quark mass one, and shed a brighter light on subtler SM effects. Recently, the LHC collider produced the definitive confirmation of abundant top quark production in $p p$ collisions; according to projections the LHC top sample by 2012 could be one order of magnitude larger than the Tevatron one. It is an exciting time for top quark physics, one where the top sector of the SM will be under serious stress.

\section*{Acknowledgments}
The author wish to thank the CDF and D0 collaborations for their effort in advancing the understanding of top quark physics, and the conference organizers for the excellent organization and the beautiful settings that stimulated exchanges between the experimental and theoretical community. In this regard, I would like to thank in particular Andreas Weiler, Jernej Kamenik, Michele Papucci, Adam Falkowski, Manuel Perez-Victoria and Gero von Gersdorff for the interesting conversations.

\end{document}